\documentclass[manuscript]{aastex}

\usepackage{tikz}
\usepackage{amsmath}

\usepackage{graphicx}

\usepackage{pdflscape}

\slugcomment{\textbf{The Astronomical Journal}}

\shorttitle{3200}
\shortauthors{Jewitt}

\begin{document}

\title{Hubble Space Telescope Observations of 3200 Phaethon At Closest Approach}


\author{David Jewitt$^{1,2}$, Max Mutchler$^3$
Jessica Agarwal$^4$, and Jing Li$^{1}$ 
} 

\affil{$^1$Department of Earth, Planetary and Space Sciences,
UCLA, 
595 Charles Young Drive East, 
Los Angeles, CA 90095-1567\\
$^2$Department of Physics and Astronomy,
UCLA, 430 Portola Plaza, Box 951547,
Los Angeles, CA 90095-1547\\
$^3$ Space Telescope Science Institute, 3700 San Martin Drive, Baltimore, MD 21218 \\
$^4$ Max Planck Institute for Solar System Research, Justus-von-Liebig-Weg 3, 37077 G\"ottingen, Germany\\
}

\email{jewitt@ucla.edu}

\begin{abstract}
We present Hubble Space Telescope observations of the active asteroid (and Geminid stream parent) 3200 Phaethon when at its closest approach to Earth (separation 0.07 AU) in 2017 December.  Images were recorded within $\sim$1\degr~of the orbital plane, providing extra sensitivity to low surface brightness caused by scattering from a large-particle trail.  We placed an upper limit to the apparent surface brightness of such a trail at 27.2 magnitudes arcsecond$^{-2}$, corresponding to an in-plane optical depth $\le 3\times10^{-9}$.  No co-moving sources brighter than absolute magnitude 26.3, corresponding to circular equivalent radius $\sim$12 m (albedo 0.12 assumed),  were detected.   Phaethon is too hot for near-surface ice to survive.  We briefly consider the thermodynamic stability of deeply-buried ice, finding that its survival would require either a very small (regolith-like) thermal diffusivity ($< 10^{-8}$ m$^2$ s$^{-1}$), or the unexpectedly recent injection of Phaethon (timescale $\lesssim$ 10$^6$ yr) into its present orbit, or both.

\end{abstract}


\section{INTRODUCTION}

Object 3200 Phaethon is one of about two dozen known ``active asteroids'', objects which have asteroid-like orbits but which show transient, comet-like morphologies (comae and/or tails) caused by mass loss (Jewitt et al.~2015a).  Mechanisms driving the mass loss in these bodies are many and varied, ranging from rotational instability and breakup, to impact, to the sublimation of recently exposed near-surface ice (Hsieh and Jewitt 2006).   In Phaethon, on-going mass loss has been detected only near perihelion, at $q$ = 0.14 AU, where it is interpreted as due to thermal fracture and/or  stresses induced by mineral desiccation at the high surface temperatures ($T \sim$ 1000 K) (Jewitt and Li 2010, Li and Jewitt 2013, Jewitt et al.~2013, Hui and Li 2017).   Mass loss rates inferred from the near-perihelion data are $dM/dt \sim 3$ kg s$^{-1}$ and the ejected particles are of micron size (Jewitt et al.~2013).   Optical observations of Phaethon against dark sky (near $r_H$ = 1 AU) have shown no evidence for dust (Hsieh and Jewitt 2005). \\

Phaethon is also the likely parent of the Geminid meteoroid stream (Williams and Wu 1993), whose mass and dynamical age are estimated as 
$M_G \sim$ (2 to 7) $\times 10^{13}$  kg (Blaauw 2017) and  $\tau_G \sim$10$^3$ years (Williams and Wu 1993, Ryabova 1999, Beech 2002, Jakub{\'{\i}}k and Neslu{\v s}an 2015), respectively.   The ratio of these quantities gives an estimate of the steady-state production rate that would be needed to supply the Geminids, namely $M_G/\tau_G \sim 7\times10^2$ to $2\times10^3$ kg s$^{-1}$.   These rates are 10$^2$ to 10$^3$ times larger than the $\sim$3 kg s$^{-1}$ estimated from the detection of a dust tail in near-perihelion data.  Indeed, hopes that the observed near-perihelion activity might account for the steady-state production of the Geminids are further dashed by the realization that the ejected (micron-sized) particles are individually 10$^3$ times smaller and 10$^9$ times less massive than the typical (millimeter-sized) Geminid meteoroids (Jewitt and Li 2010).  Such small particles are strongly accelerated by solar radiation pressure and do not contribute to the Geminids stream.  Debris particles with sizes $\gtrsim$ 10 $\mu$m have been reported from infrared observations (Arendt 2014) and recently, even larger  Geminids (up to 2 cm) were inferred via their impact ejection of dust from the Moon (Szalay et al.~2018).  However, instead of being a steady-state phenomenon, it appears more likely that the production of the Geminid stream is associated with an unspecified transient or even impulsive  event occurring within the last few $\sim$10$^3$ yr.  The nature of this event is unknown.

Phaethon's orbit  has semimajor axis $a$ = 1.271 AU, $e$ = eccentricity 0.890 and  inclination $i$ = 22\degr.  The resulting Tisserand parameter with respect to Jupiter is $T_J$ = 4.509.  This large value is inconsistent with Phaethon being a captured Kuiper belt (2 $\le T_J \le$ 3) or Oort cloud ($T_J <$ 2) comet (however, see Ryabova 2007 for a contrary view) but is typical of the asteroids (for which $T_J >$ 3).   Indeed, there exists a  purported dynamical pathway linking Phaethon to the 500 km diameter  main-belt asteroid 2 Pallas (de Leon et al.~2010, Todorovi{\'c} 2018), raising the suspicion that the former could be a fragment of the latter. While the reflection spectra of Phaethon and Pallas do show significant differences, both are blue relative to the Sun (i.e.~they are asteroid spectral B-types), broadly consistent with a physical connection.  Two other asteroids appear dynamically associated with Phaethon; 2005 UD (Ohtsuka et al.~2006) is another B-type (Jewitt and Hsieh 2006) while 1999 YC is a  C-type (Kasuga and Jewitt 2008), enhancing the sense of a past catastrophic event (although not  the same one producing the Geminids).  Together, these objects define the so-called ``Phaethon-Geminid Complex'' (Kasuga 2009).

Here, we present Hubble Space Telescope observations of 3200 Phaethon taken near the time of its closest approach to Earth in 2017 December, at geocentric distance $\Delta$ = 0.07 AU.  In addition, our data were taken within 1\degr~of the Earth crossing Phaethon's orbital plane, providing sensitivity to large, slow-moving particles confined to a trail.   These observations are independent of, but complementary to, HST observations taken before Phaethon at a slightly larger geocentric distance and out-of-plane angle on UT 2017 December 14 under GO 15357 (Ye et al.~2018).

\section{OBSERVATIONS}
\label{observations}
We used the Wide Field Camera 3 (WFC3) imager on the 2.4 m Hubble Space Telescope, with all observations taken under General Observer program GO 15343.  The UVIS channel of this instrument uses two thinned, backside-illuminated charge-coupled devices (CCDs) each with 2051$\times$4096 15 $\mu$m pixels, separated by a 1.2\arcsec~gap.  The  image scale is 0.04\arcsec~pixel$^{-1}$, while the Nyquist-sampled (2 pixel) resolution of the data corresponds to a linear distance of 4 km at the distance of Phaethon.  The full angular field of view of the instrument (162\arcsec$\times$162\arcsec)~corresponds to 8200 km at Phaethon.   We used the wide F350LP filter in order to maximise sensitivity to faint sources in the data.  This filter benefits from a peak system throughput of 29\% and takes in most of the optical spectrum with wavelengths $\lambda >$ 3500\AA.  The effective wavelength is 5846\AA~and the effective width 4758\AA.

We obtained images in three sets, designated A, B and C in Table (\ref{obslog}), within each of which we secured four short (4 s) integrations (labeled ``S''), three medium (30 s) integrations (``M'') and three long (348 s) integrations (``L'', Table \ref{obslog}).  These sets are separated by $\sim$1.5 hours, during which the observing geometry changed only slightly.  The most significant change was in the angle between the line of sight and Phaethon's orbital plane, which varied from $\sim$0.3\degr~to $\sim$1.2\degr~(Table \ref{geometry}).  The S, M and L integration times were chosen to maximise sensitivity to diffuse material and to objects at different distances from Phaethon, recognizing the high surface brightness of scattered light at small angular separations.  Whereas the long integrations within each set were read-out full-frame, we used sub-frame readout for the medium and short integrations in order maximize the observing efficiency (Table \ref{obslog}).  The orientation of the WFC3 field was rotated by $\sim$20\degr~between the A, B and C image sets in an attempt to identify artefacts in the data.  We assumed that instrumental and scattered light artefacts  would rotate with the telescope while real  features associated with Phaethon would instead maintain a fixed sky position angle.   In this, we emulated procedures used before to search for faint objects in the vicinity of bright targets including potential satellites of Pluto (Weaver, et al.~2006),  2 Vesta (McFadden et al.~2015), and 1 Ceres (DeMario et al.~2016).  A log of observations is given in Table (\ref{obslog}). The hour-by-hour geometry of the observations is indicated in Table (\ref{geometry}).

\begin{figure}
\epsscale{1.00}
\plotone{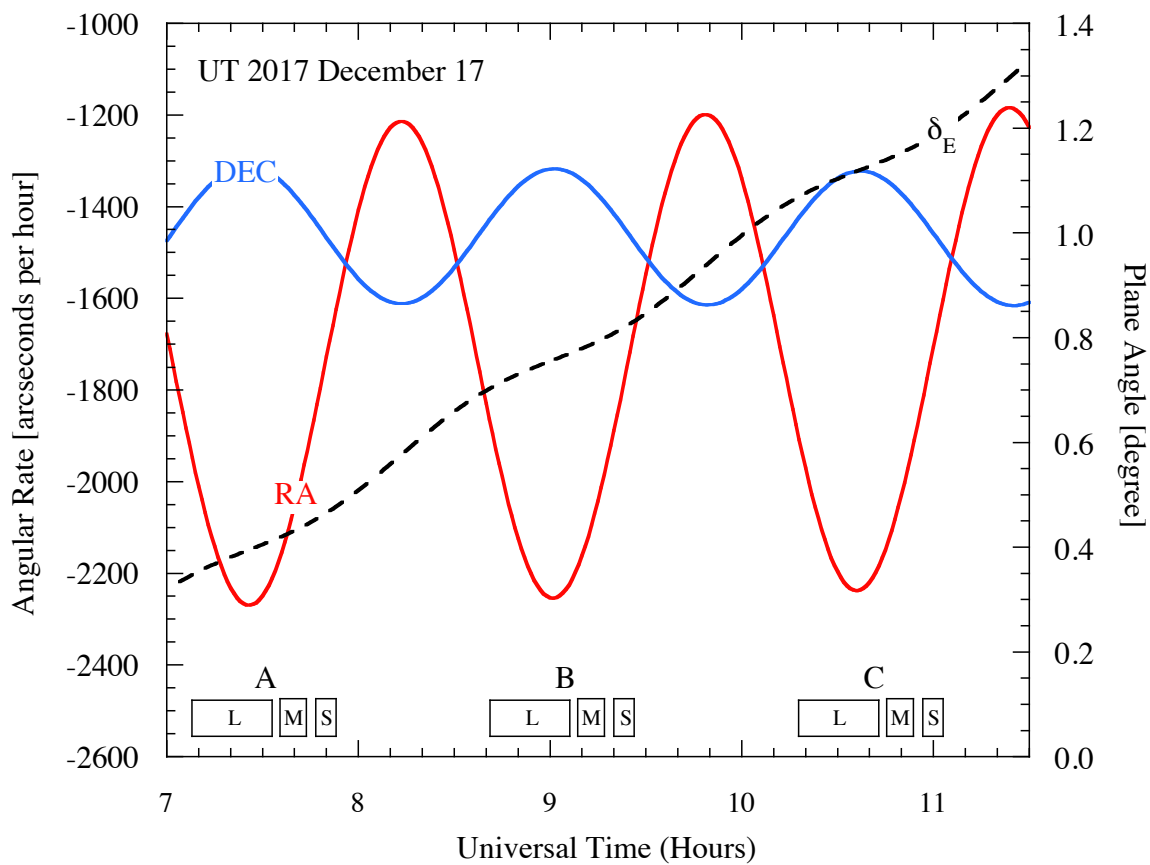}
\caption{(left axis) Angular non-sidereal rates of motion of Phaethon in right ascension (red curve labeled RA) and declination (blue curve labeled DEC) from HST on UT 2017 December 17; (right axis) Out of plane angle (black, dashed curve labeled $\delta_E$).  Labeled boxes at the bottom indicate the times of the long (L), medium (M) and short (S) integrations in each of the three sets, A, B and C (see Table \ref{obslog}).
\label{rates}}
\end{figure}

During the observations Phaethon moved at angular rates up to $\sim$2300\arcsec~hr$^{-1}$ in right ascension and 1600\arcsec~hr$^{-1}$ in declination (Table \ref{geometry}, Figure \ref{rates}), relative to field stars and galaxies, as a result of parallax caused by the small HST - Phaethon distance.  This is equivalent to trailing of field objects by up to 18 WFC3 pixels s$^{-1}$.   The telescope uses linear interpolation between ephemeris positions that leads to trailing  of Phaethon in the longer ``L'' integrations and, to a lesser extent, in the medium (``M'') integrations.  Trailing is not an issue in the short ``S'' integrations.

Figure (\ref{image_long})  shows a single WFC3 image annotated to mark some of the artefacts of note as well as the cardinal directions and a scale bar. In the Figure, $a a'$ marks the 1.2\arcsec~wide inter-chip gap, the secondary mirror support diffraction spikes are labeled $b b'$ and $c c'$, while $d d'$ marks charge bleed from the saturated image of the nucleus.  Diffuse objects marked $e1$ to $e4$ are  out-of-focus internal reflections from the bright field object in the lower right while $f f'$ is internally scattered light from an unknown field source.  Diffuse doughnut ``g" is an out-of-focus internal reflection of Phaethon.  Not marked in the Figure are huge numbers of ``cosmic ray'' (charged particle) tracks and  numerous trailed field objects. 

Other than the doughnut feature ``g'', the internal reflections vary in position, shape and brightness from image to image and are effectively removed in median combinations of the data. The cosmic ray tracks are likewise easily eliminated.  The trailed field stars and galaxies are so numerous and so extended, however, that they cannot be completely removed from the data.  They set the ultimate limits to the sensitivity in our search for near-Phaethon material.  In some cases, the medians of image sets A, B and C show point-like residuals caused by the overlap of different trails.  These are identified as artefacts, however, because they do not show the correlated motion between  image sets expected of a real object.  

\section{DISCUSSION}

No Phaethon-associated dust  is evident in our data and so we seek to place quantitative upper limits to the presence of such material.  The spatially complex and temporally variable  background to the Phaethon HST data makes it impossible to assign a single value to the limiting sensitivity.  In addition, the sensivity to diffuse emission, as would be expected from near nucleus dust, depends on many factors including both the surface brightness and the solid angle subtended.  The morphology is unknown and unknowable because it is a function of the ejection mechanism, its particle size and angular dependence and the subsequent motion of ejected particles under the action of solar radiation pressure.   Our approach is to set the most conservative limits, paying particular attention to dust at the anticipated close-approach distance of JAXA's planned  DESTINY+ mission to Phaethon (see section \ref{destiny}).

\subsection{Surface Brightness Limits: Dust Trail}

Large, slow-moving particles ejected from a parent body stay close to the orbital plane of the parent, forming a line-like ``trail'' on the sky when observations are taken from a vantage point near the orbital plane.  The surface brightness of such a trail is geometrically enhanced, potentially by a large factor.   Our HST images were targeted at small out-of-plane angles, 0.3\degr~$\le \delta_{\oplus} \le$ 1.2\degr~(c.f.~Table \ref{geometry}), giving us high sensitivity to a large particle dust trail.

The surface brightness of a uniform source is independent of the distance to the observer, $\Delta$, but should vary with heliocentric distance, $r_H$, and phase angle, $\alpha$.  The phase angle dependence is described by the phase function, $\Phi(\alpha) \le 1$, equal to the ratio of the light scattered at angle $\alpha$ to that at $\alpha$ = 0\degr.  The HST observations were taken at large phase angle, 70.0\degr~$\le \alpha \le$ 72.5\degr~(Table \ref{geometry}), and so the phase correction deserves comment.  

The relevant phase function for Phaethon dust is unclear, because the properties (indeed, the existence) of the dust are unknown.  Optically large dust particles (i.e.~those with radius $a \gg \lambda$, where $\lambda$ is the wavelength of observation), of the type expected to be confined to a trail, might possess phase functions similar to those of asteroids. In this regard, the S- and C-types have $\Phi(71\degr)$ =  0.12 and 0.10, respectively, as computed from the Bowell et al~(1989) formalism with angle parameters $g$ = 0.25 and 0.15, respectively.  Empirically, the phase functions   for dusty comets tend to be strongly forward-scattering but otherwise relatively flat.  For example, the comet dust phase function of Marcus (2007) gives $\Phi(71\degr) \sim$ 0.4.  In the following we adopt the largest phase correction, $\Phi(71\degr)$ = 0.10, corresponding to $2.5\log_{10}(\Phi(71\degr))$ =  -2.5 magnitudes, in order to maximize the possible surface brightness (hence cross-section) of dust at opposition and so to obtain the most robust upper limits to the possible dust optical depth.

We interpret the light  from 1 square arcsecond as being scattered from dust of cross section, $C_1$ (m$^2$), given by the inverse square law  (c.f.~Russell 1916) 

\begin{equation}
C_1 = 2.25\times10^{22} \pi \left(\frac{ r_H^2 \Delta^2}{p_v \Phi(\alpha)}\right) 10^{0.4 \Delta V}
\label{invsq}
\end{equation}

\noindent where $r_H$ and $\Delta$ are expressed in AU and $p_v$ is the geometric V-band albedo.  We adopt $p_V$ = 0.12$\pm$0.01 as reported for the nucleus of Phaethon by Hanu{\v{s}} et al.~(2016), while noting that the albedo of the dust could be higher or lower, depending on the scattering properties of the grains.   Quantity $\Delta V = V_{\odot} - V_1$, where $V_{\odot} = -26.77$ is the magnitude of the Sun and $V_1$ is the magnitude of a single square arcsecond of the dust, which is numerically equal to the surface brightness.  The optical depth is given by $\tau = C_1/s^2$, where $s = 7.25\times10^5 \Delta$ is the linear distance at Phaethon, in meters, corresponding to 1 arcsecond.  We obtain

\begin{equation}
\tau = 1.3\times10^{11} \left(\frac{ r_H^2}{p_v \Phi(\alpha)}\right) 10^{0.4 \Delta V}
\label{tau}
\end{equation}

\noindent provided $\tau <$ 1.

\begin{figure}
\epsscale{0.85}
\plotone{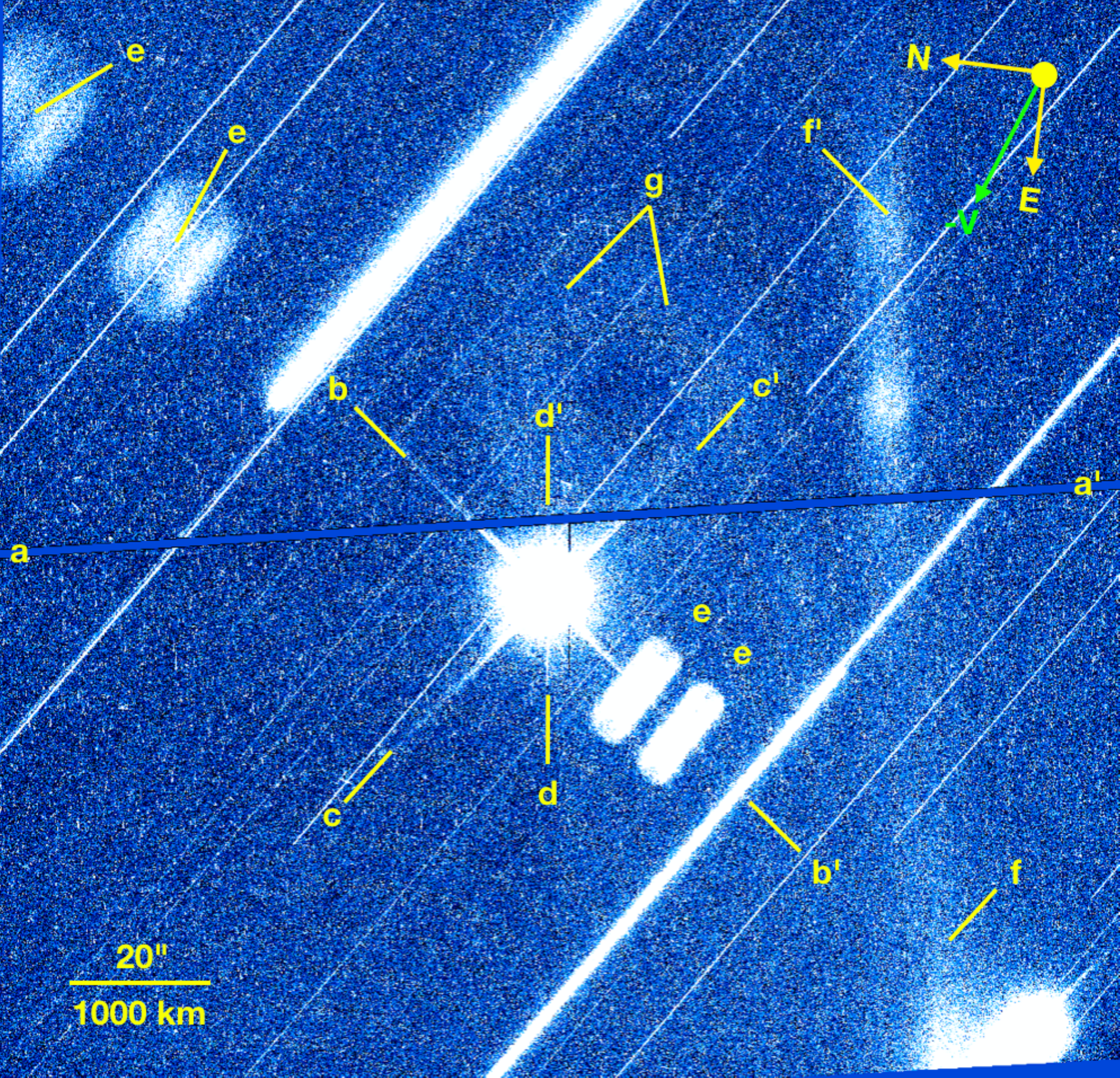}
\caption{Single, unprocessed 348 s WFC3 image of Phaethon with image artefacts marked (see \textsection \ref{observations}).  $N$ and $E$ show the cardinal directions while, in the lower left, a 20\arcsec~scale bar corresponds to 1000 km at the $\Delta$ = 0.07 AU distance to Phaethon.  The green arrow shows the (overlapping) directions of the negative heliocentric velocity vector and the anti-solar direction (see Table \ref{geometry}).  
\label{image_long}}
\end{figure}

In our data, the near-Phaethon space is occupied by numerous trailed field stars and galaxies, as well as by scattered light artefacts and numerous cosmic rays, as already discussed (Figure \ref{image_long}).  The cosmic rays were easily eliminated by forming combinations of the three images with a given exposure taken within each set.   However, attempts to self-subtract the artefacts by using image combinations taken with different telescope roll angles were largely unproductive.  This is because the background sources of internally scattered light changed from image-to-image as HST followed the rapid motion of Phaethon across the sky.  As a result, the internal reflections move and change in intensity in a way that is too complicated for roll subtraction to correct.  Simply put, no two images look the same.  Still we find that taking the data with different roll angles was useful in helping us to search for dust associated with Phaethon, because such material must necessarily hold a fixed sky-plane position angle on the timescale of these observations.  No such dust was found.

The surface brightness of the artefacts provides an immediate upper limit to the possible surface brightness of diffuse sources near Phaethon.  For example, the brightest internal reflection, labeled $e3$ in the long (348 s) image composites (Figure \ref{image_long}), has peak V-band surface brightness $\Sigma(r_H,\alpha)$ = 25.5 magnitudes arcsec$^{-2}$, while feature $g$ has $\Sigma(r_H,\alpha)$ = 26.4 magnitudes arcsec$^{-2}$ and fainter diffuse sources can readily be detected over most of the image plane.  A first-order, conservative conclusion from examination of the images is that no extended sources with surface brightness $\Sigma(r_H,\alpha) \le$ 26 magnitudes arcsecond$^{-2}$ exist in the long-exposure images of Phaethon beyond separation angles $\theta >$ 10\arcsec~to 15\arcsec.  The surface brightness limit decreases closer to Phaethon, and becomes a complicated function of radius and azimuth near to the saturated image core.    Substitution into Equation (\ref{tau}) then gives  $\tau < $ 9$\times$10$^{-9}$ as a conservative upper limit to the line-of-sight optical depth in dust.

A stronger constraint can be obtained by considering the likely distribution of Phaethon dust.  Observations of comets and active asteroids from vantage points close to their orbital planes often show dust trails, caused by large dust particles ejected slowly from their source bodies.  Radiation pressure plays no role in setting the thicknesses of the trails because it acts only in the radial direction from the Sun.  Instead, the trail thickness is determined primarily by  the component of the ejection velocity measured perpendicular to the orbital plane.  Such trails can be extraordinarily thin because the dust ejection velocities are remarkably small, typically being comparable to the gravitational escape speed from the parent body.  For example, active asteroid 133P/Elst-Pizarro ejects dust at $\sim$1 to 2 m s$^{-1}$ and has a trail with full width at half maximum (FWHM) = 300 km at 1000 km from the nucleus (Jewitt et al.~2014).  Active asteroid 311P, when viewed from within the orbit plane, shows a large-particle trail with FWHM rising from 400 km near the nucleus to 600 km some 30,000 km from it (Jewitt et al.~2015b).  The derived ejection velocity from 311P is $<$ 1 m s$^{-1}$.   Accordingly, we searched the HST data for evidence of a narrow dust trail in Phaethon, using the morphological evidence from other active asteroids as a model.  

\begin{figure}
\epsscale{1.00}
\plotone{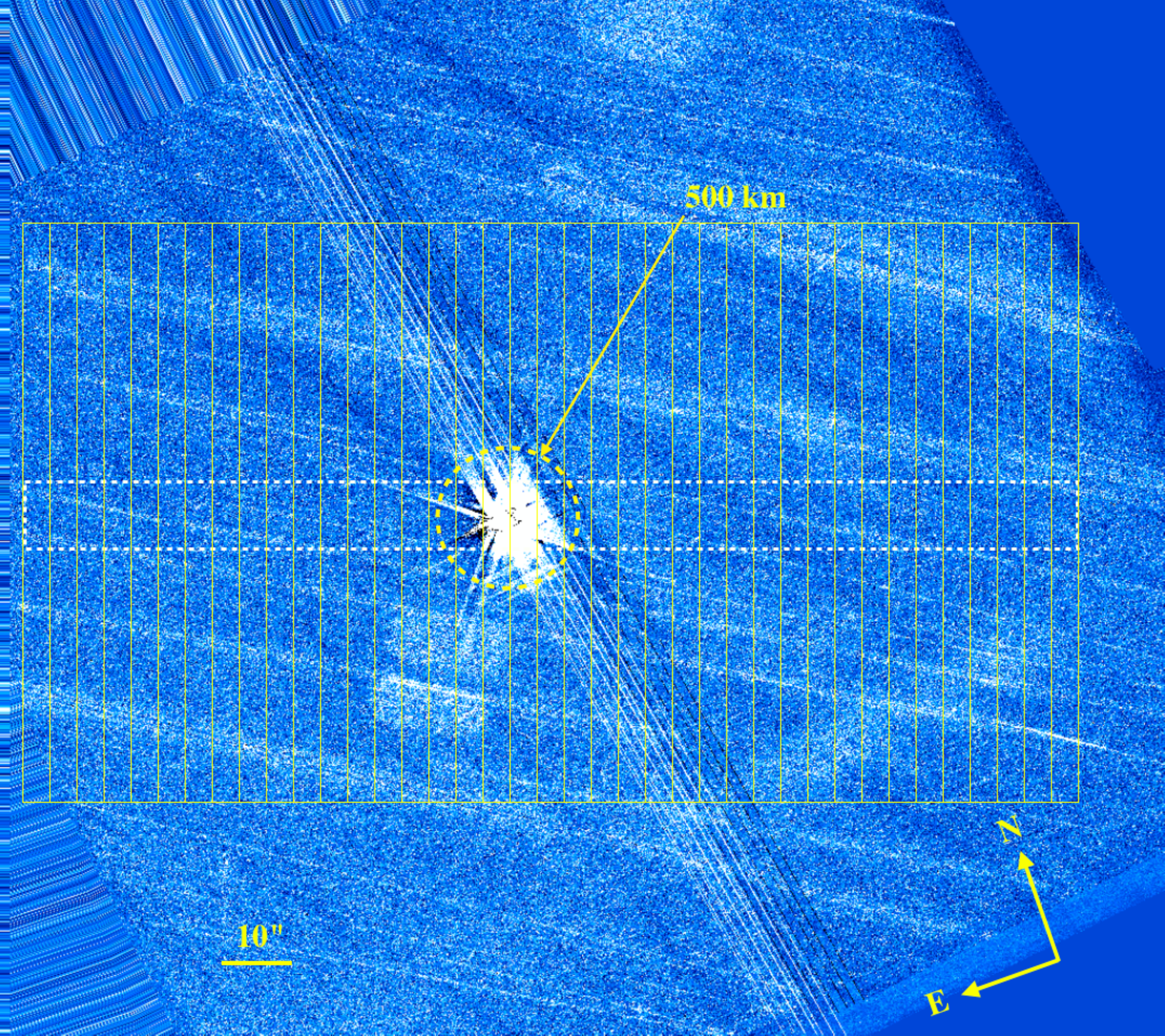}
\caption{Cosmic-ray rejected long-exposure (total integration 1044 s) image showing the extraction strips used to compute surface brightness profiles perpendicular to the projected orbit, which has been rotated to the horizontal in this image.  The extraction boxes, shown in thin yellow lines, are 4\arcsec$\times$80\arcsec~in size.  The apparent width of a 500 km dust trail is marked by a dotted white box.  The anticipated 500 km close-approach distance of DESTINY+ is marked as a yellow, dashed circle, for reference.   No dust trail is evident.  
\label{cuts}}
\end{figure}

\begin{figure}
\epsscale{1.00}
\plotone{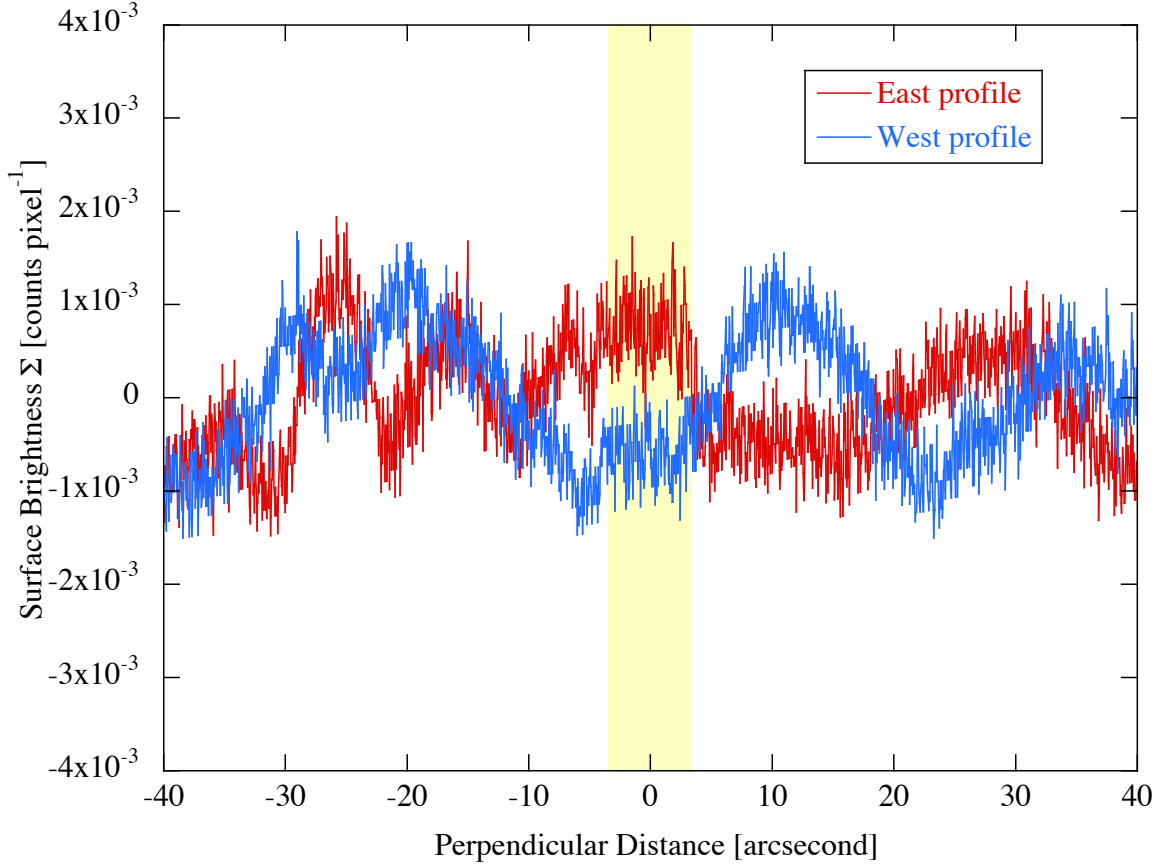}
\caption{Sample median surface brightness cuts perpendicular to the projected orbital plane.  Red and blue lines show cuts measured east and west of the nucleus.  The yellow band shows the expected location and approximate width of the in-plane dust trail.   Structure in the profiles is the result of low surface brightness background sources in the data.  Surface brightness is expressed in instrumental units, where 1 count pixel$^{-1}$ corresponds to $\Sigma(r_H, \alpha)$ = 19.7 magnitude arcsec$^{-2}$.  Values are expressed relative to the mean across the image.
\label{perp_cuts}}
\end{figure}

At $\Delta$ = 0.07 AU, a 500 km FWHM trail would subtend $\sim$10\arcsec.  Figure (\ref{cuts}) shows a sample long-integration image of Phaethon formed by combining images A-L (1-3) to remove cosmic rays and rotated to bring the projected orbit direction to the horizontal.  Overlaid on the image (in dashed white lines) is a box of projected width 10\arcsec.  We computed a set of surface brightness profiles perpendicular to the expected trail direction (i.e.~along the thin vertical boxes shown in the Figure).  Each profile sampled 4\arcsec~along the horizontal direction and extended $\pm$40\arcsec~above and below it.  We then computed the median of these vertical profiles, in order to suppress trailed field objects and other artefacts which appear at different positions relative to the projected orbit.  Example median profiles are shown in Figure (\ref{perp_cuts}), where a linear gradient has been subtracted from the data and the residual bumps in the profile are due to imperfect removal of background sources.  A large particle trail in Figure (\ref{perp_cuts}) would appear as a symmetric excess at $x$ = 0\arcsec~and 10\arcsec~wide if FWHM = 500 km.   No consistent excess is apparent  in the profiles shown, neither do other profiles or combinations of profiles computed from the HST images show any hint of an orbit-aligned trail, of any thickness.  It is clear from the Figure that systematic uncertainties (caused by imperfectly removed field objects) are larger than statistical errors, and so it is difficult to assign a formal 3$\sigma$ limit to the surface brightness of dust.  By comparing different combinations of vertical cuts, we conclude that a firm upper limit to the surface brightness of any orbit-aligned, 133P-like trail with a perpendicular scale $\sim$10\arcsec~(500 km) is  10$^{-3}$ counts per pixel, corresponding to $\Sigma(r_H,\alpha) \ge$ 27.2 magnitudes arcsecond$^{-2}$.   By Equation (\ref{tau}) with $V_1 \ge$ 27.2, we find that this corresponds to optical depth $\tau \le 3\times 10^{-9}$.  This is comparable to trail optical depth limits (10$^{-8}$ to 10$^{-9}$) set using ground-based data (Ishiguro et al.~2009) but probes the dust environment much closer to Phaethon than was previously possible.

\subsection{Point Sources}

\begin{figure}
\epsscale{0.85}
\plotone{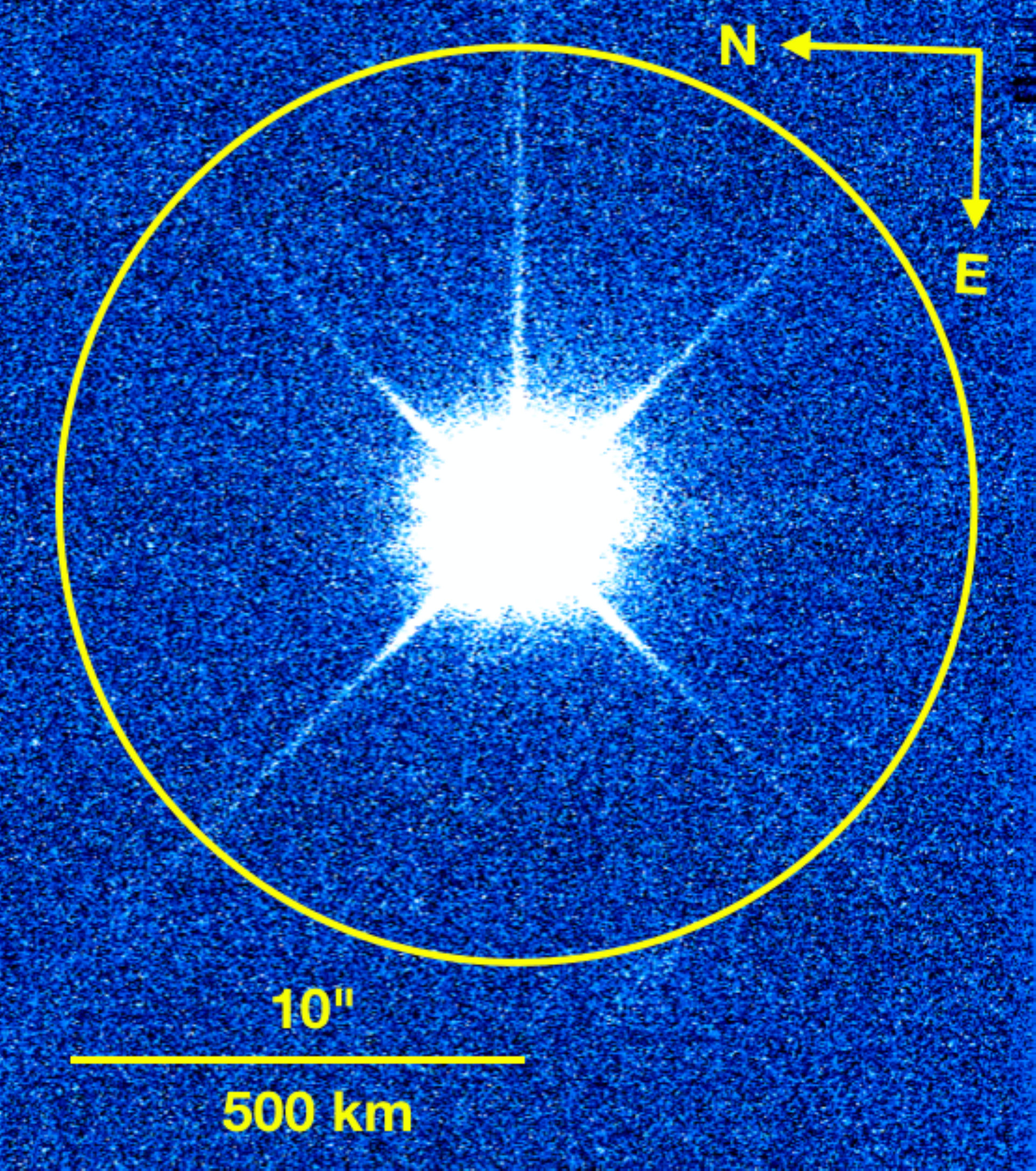}
\caption{Example composite of 4$\times$4 s WFC3 integrations (see \textsection \ref{observations}).  $N$ and $E$ show the cardinal directions while a scale bar is given in the lower left.  The diffuse feature extending to the south of Phaethon (position angle $\sim$195\degr), like the elliptical object 10\arcsec~to its east (105\degr), is a residual artefact not seen in the other short exposure composites.  The radius of the circle shows the anticipated 500 km DESTINY+ close-approach distance.
\label{image_short}}
\end{figure}

We use only the short exposure images to set limits to the brightness of possible co-moving point sources, caused by boulders that might have been ejected from Phaethon.  The short exposure images do not suffer from guiding drift and are affected by far fewer trailed field objects than the long exposure images.  An example is shown in Figure (\ref{image_short}).  No co-moving point sources are apparent in the HST data.  We set a limit to the brightness of such objects by adding artifical stars to the data to find that point sources with $V \le$ 23 are detectable over most of the field of view. This limit decreases very close to Phaethon ($\theta \lesssim$ 5\arcsec) because of scattered light, but is valid at 10\arcsec~(500 km) separation (see yellow circle in the Figure).  At $r_H$ = 1.004 AU, $\Delta$ = 0.069 AU and $\alpha$ = 71\degr, $V \ge$ 23.0 corresponds to absolute magnitude $H_V \ge$ 26.3.  With $p_V$ = 0.12, this absolute magnitude sets a limit to Phaethon co-moving companion radii $a \le$ 12 m.  

The Hill sphere of Phaethon has a radius of about 78 km (subtending $\theta$ = 1.5\arcsec~at 0.07 AU).  Even in our shortest integrations, the near-nucleus region is obliterated by saturated pixels (for $\theta \le$ 0.7\arcsec) or strongly contaminated by scattered light, and so cannot be meaningfully studied.  Separate observations obtained in the thermal infrared were secured to better study the Hill sphere and will be presented in another paper.

\subsection{DESTINY+}
\label{destiny}
The planned Japan Aerospace Exploration Agency (``JAXA'')  mission to Phaethon, called ``DESTINY+'' (Arai et al.~2018), has an intended close-approach distance of 500 km (Ishibashi et al.~2018).  This corresponds  to an angular distance $\theta$ = 10\arcsec~at $\Delta$ = 0.07 AU, which falls well within the field of view of the present dataset (see Figure \ref{cuts}).  We briefly consider some  consequences of the non-detection of dust for the DESTINY+ mission.

The DESTINY+ spacecraft is box-like with a longitudinal cross-section $A_s$ = 0.9 m$^2$ (Arai et al.~2018).   The total cross section of all the particles to be intercepted in this area is just $C_s = A_s \tau$.  Substituting for $\tau$, we find $C_s \le $  3$\times$10$^{-9}$ m$^2$.  This is equal to the  cross-section of a single, spherical particle having radius $a = (A_s \tau/\pi)^{1/2}$.   We find $a$ = 30 $\mu$m, which is much smaller than the characteristic millimeter size of the Geminid meteoroids (Schult et al.~2018), leading us to conclude that it is unlikely that DESTINY+ will hit such a particle.    The corresponding particle mass is $M = 4/3 \pi \rho (A_s \tau/\pi)^{3/2}$.  With density $\rho$ = 2500 kg m$^{-3}$, we find $M$ = 3$\times$10$^{-10}$ kg.  Any  particle intercepted by the spacecraft would likely be small and a product of recent activity; the travel time to reach 500 km at 3 m s$^{-1}$ is only $\sim$2 days.

DESTINY+ will carry a dust analyzer with collecting area $A_d = 0.011$ m$^2$ (Masanori et al.~2018).  If flying in the orbital plane of Phaethon, the collector will capture dust with a total cross-section ($A_d \tau <$ 3$\times$10$^{-11}$ m$^2$), equivalent to a single spherical particle of mass $M < 3\times$10$^{-13}$ kg and radius $a <$ 3 $\mu$m.   Of course, this tiny intercepted cross-section and mass will be distributed over a  number of particles, depending upon the size distribution and size range of the particles released from Phaethon.    Near-perihelion observations indicate that micron-sized dust particles are released from Phaethon at rates $dm/dt \sim$ 3 kg s$^{-1}$, sustained for $\sim$1 day (Jewitt et al.~2013).   Assuming, for the sake argument, that the particles are released isotropically  at speed, $U$, their number density at distance $L$ from the nucleus would be 

\begin{equation}
N_1 = \frac{1}{4\pi U L^2} \left(\frac{3}{4 \pi \rho a^3 }\right) \frac{dm}{dt}
\end{equation} 

\noindent and the resulting optical depth along a line of sight having impact parameter, $L$ is 

\begin{equation}
\tau \sim N_1 \pi a^2 L.
\end{equation}

\noindent to within a factor of order unity.  Substituting $dm/dt$ = 3 kg s$^{-1}$, $\rho$ = 2500 kg m$^{-3}$, $a$ = 10$^{-6}$ m, $U$ = 3 m s$^{-1}$ (i.e.~roughly the escape speed from Phaethon) and $L$ = 5$\times$10$^5$ m, we find $N_1$ = 30 m$^{-3}$ and $\tau \sim$ 5$\times$10$^{-5}$, which is about four orders of magnitude larger than the upper limit to the optical depth we have placed.  While this calculation is necessarily very crude, we can  conclude with confidence that a coma of micron-sized grains released at near-perihelion rates would have been easily detected in our data.  An optical depth limit $\tau < 3\times10^{-9}$ restricts the possible release of such grains at $r_H$ = 1 AU to physically insignificant rates $dm/dt \lesssim$ 3$\times$10$^{-4}$  kg s$^{-1}$.
%

The Geminids are more nearly millimeter-sized rather than micron-sized.  Measurements in the $10^{-7} \le m \le 10^{-2}$ kg mass range (corresponding radii $0.2 \le a \le 14$ mm, with $\rho$ = 2500 kg m$^{-3}$) support a differential power law distribution, $dN \propto m^{-s} dm$ with mass index $s$ = 1.68$\pm$0.04 (Blaauw 2017, c.f.~1.64 by Schult et al.~2018).  This mass index corresponds to a differential size distribution index $q = 3s - 2 = 3.04\pm$0.04.  In such a distribution, the cross-section is spread approximately uniformly in equal increments of log(a) while the mass is dominated by the largest particles.
Particles smaller than 0.2 mm and larger than 14 mm  no doubt exist outside the range sampled by Blaauw (2017). However, Geminids smaller than about $a_0 \lesssim$ 10 $\mu$m are quickly swept from the stream by radiation pressure and should be absent or severely depleted unless recently released from the nucleus.  Larger particles, up to $a_1 \sim$ 2 cm, have been inferred from dust detectors orbiting the Moon (Szalay et al.~2018) and up to 4 cm from bolides (Madiedo et al.~2013). 

We give an example for  the case of the optical detection of larger particles. The number density corresponding to optical depth $\tau$ is $N_1 = \tau/(\pi a_o^2 \ell)$, where we assume an average radius $a_0$ and $\ell$ is the path length along the line of sight.  We take $\ell \sim L$, with $L$ = 500 km as the minimum approach distance of the spacecraft and $a_0$ = 0.5 mm.  Then, the number density of particles is $N_1 < 7\times 10^{-9}$ m$^3$ and their average separation is $N_1^{-1/3} \gtrsim$ 500 m.  Any chance that these particles could be directly imaged  (as were nearby dust particles in the ROSETTA rendezvous with comet 67P/Churuyumov-Gerasimenko, see Ott et al.~2017) are dashed by the extreme parallactic trailing as the fast-moving spacecraft tracks the nucleus of Phaethon.  Dust particles in the vicinity of the nucleus will not suffer from this trailing but will be too distant and faint to be detected.

\subsection{Mechanisms}
The question of how an asteroid can eject millimeter-sized (and probably larger) rocks to populate the Geminid meteoroid stream has existed since the discovery of Phaethon in 1983.  In comets, mass loss is driven by drag forces from expanding gas generated by the sublimation of near-surface ice.  Some main-belt asteroids (the ``main-belt comets'') appear to contain ice (Hsieh and Jewitt 2006) and so we consider this possibility first.  Internal ice could exist if Phaethon originated in an outer-belt orbit, for example as a fragment of 2 Pallas, or in a more distant location.  Can ice survive in Phaethon? 

It is easy to reject the possibility of near-surface ice, because the surface temperature at perihelion is so high ($T \sim$ 1000 K); surface ice would be long-gone.  The survival of deeply buried ice (although it could not drive the near-perihelion activity) is more difficult to address, requiring the calculation of a detailed thermal model of the object coupled with a model of the dynamical evolution.  The thermophysical and dynamical evolutionary parameters of Phaethon are not sufficiently well known to justify such a calculation. Instead, we proceed as follows.  

The e-folding timescale  for heat to conduct from the surface to the core of Phaethon is $\tau_c \sim r_n^2/\kappa$, where $r_n$ = 3 km is the nucleus radius and $\kappa$ is the thermal diffusivity of the material of which Phaethon is made.  On timescales $\gg \tau_c$ the core temperature, $T_c$, should approach the temperature of a spherical blackbody in equilibrium with the solar insolation averaged around its orbit.  A calculation gives $T_c$ = 300 K, which is far too high for ice to exist (Jewitt and Li 2010).

However, if it were initially present, ice could survive in the core of  Phaethon provided $\tau_c > \tau_o$, where $\tau_o$ is the time elapsed since  entering the present, small-perihelion orbit.  For this inequality to be true, the bulk diffusivity of Phaethon must be $\kappa < r_n^2/\tau_o$.   The timescale $\tau_o$ is not known, but an indication is provided by numerical integrations of the orbital evolution, which give a half life of $\tau_o$ = 26 Myr (de Le{\'o}n et al.~2010).   Substitution of this value gives $\kappa < 10^{-8}$ m$^{2}$ s$^{-1}$, which is about two orders of magnitude smaller than the diffusivities typical of dielectric solids  ($\kappa \sim$ 10$^{-6}$ m$^2$ s$^{-1}$).  Jakub{\'{\i}}k \& Neslu{\v s}an (2015) report dynamical stability on an even longer timescale ($\sim$100 Myr, implying $\kappa < 3\times 10^{-9}$ m$^2$ s$^{-1}$).  Such low diffusivities are  found in highly porous, regolith type materials (e.g.~the diffusivity of the lunar regolith is $5\times10^{-9} \le \kappa \le 5\times10^{-8}$ m$^2$ s$^{-1}$ according to parameters given by Rubanenko and Aharonson 2017).  However, high porosities are difficult to reconcile with the robust nature of the Geminid meteoroids, as inferred from their deep penetration  into the Earth's atmosphere.  Specifically, Geminid densities ($\rho$ = 2900$\pm$600 kg m$^{-3}$, Babadzhanov and Kokhirova 2009) and tensile strengths ($\sim 3\times10^5$ N m$^{-2}$, Beech et al.~2002) are amongst the highest of any measured values and seemingly at odds with a porous structure.  Only if $\tau_o \ll$ 26 Myr could the conduction time be long enough to preserve buried ice without requiring regolith-like porosity.  For example, if $\tau_o$ = 1 Myr, then we require only that $\kappa < 0.3 \times 10^{-6}$ m$^2$ s$^{-1}$, much more compatible with solid rock.   Simply put, while we cannot empirically rule out the possibility that Phaethon retains deeply buried ice, its survival would require either an unexpectedly small thermal diffusivity or an unexpectedly recent deflection of Phaethon into its present orbit from a more distant one.


As noted above, thermal fracture and/or the build-up of desiccation stresses are probably responsible for the near-perihelion activity.  These processes are likely assisted by radiation pressure sweeping which, at perihelion, can remove sub-millimeter particles once they are detached from the surface (c.f.~Equation (17) of Jewitt 2012).  The very high linear polarization ($\sim$50\% at phase angle 100\degr) exhibited by Phaethon has been interpreted as evidence for the preferential loss of small particles through this effect (Ito et al.~2018).  Rotational instability could play a contributory role.  The  rotation period of 3.6 hr (Ansdell et al.~2014, Hanu{\v{s}} et al.~2016)  is not remarkable compared to asteroids of comparable size and the rotational frequency (6.7 day$^{-1}$) is significantly below the empirical rotational barrier (11 day$^{-1}$)   (Pravec et al.~2002).   However, depending on the shape of Phaethon, the centripetal acceleration near the equator could be a large fraction of the gravitational acceleration there, so raising the maximum size of particles that can be removed by radiation pressure sweeping.  Indeed, some lightcurve-based shape models (e.g.~the top panel of Figure 5 in Hanus et al.~2016) hint at a muffin-top shape consistent with the equatorward movement of surface material owing to incipient rotational instability.  Rotational instability in small asteroids and comets is a natural consequence of both the YORP effect and, when it is present, of outgassing. 

Thus, we envision that some combination of thermal or desiccation stresses, fast rotation and radiation pressure sweeping is responsible for the mass loss detected near perihelion.  However, the apparent lack of activity away from perihelion leaves the origin of the  Geminid meteoroid stream unresolved.

\clearpage

\section{SUMMARY}

We describe Hubble Space Telescope observations of active asteroid (and Geminid parent) 3200 Phaethon, taken within $\sim$1\degr~of orbit plane crossing and at the closest approach to the Earth (geocentric distance $\Delta$ = 0.069 AU) in 2017 December.  

\begin{enumerate}

\item A limit to the apparent surface brightness of a large-particle dust trail is set at $\Sigma(r_H,\alpha) \ge$ 27.2 magnitudes arcsec$^{-2}$.  After correction for phase darkening, we find a limit to the optical depth of dust in the orbit plane $\tau \le 3\times10^{-9}$.

\item Assuming that the activity level of Phaethon remains fixed, the corresponding upper limit to the mass which will be intercepted by the DESTINY+ dust detector, when passing Phaethon at 500 km minimum distance, is comparable to that of a single spherical particle 3 $\mu$m in radius.

\item The observations provide no evidence for comoving point sources brighter than $V \sim$ 23 (spherical equivalent radius 12 m, at assumed geometric albedo 0.12).

\item Deeply buried ice could survive in Phaethon over the estimated 26 Myr dynamical lifetime only if the thermal diffusivity is regolith-like ($\kappa < 10^{-8}$ m$^2$ s$^{-1}$), but such low diffusivities are difficult to reconcile with the rugged mechanical properties of the Geminid meteors.  
\end{enumerate}

\acknowledgments
We thank Quan-Zhi Ye, Toshi Kasuga, Pedro Lacerda, Yang Bin and Tomoko Arai for reading the manuscript, Ned Wright for pointing out an important error in an earlier version of this manuscript and the referee, G. Ryabova, for a review.  Based on observations made under GO 15343 with the NASA/ESA Hubble Space Telescope, obtained at the Space Telescope Science Institute,  operated by the Association of Universities for Research in Astronomy, Inc., under NASA contract NAS 5-26555.



{\it Facilities:}  \facility{HST}.

\clearpage


\clearpage

\begin{deluxetable}{llcrcrccccr}
\tablecaption{Observing Log 
\label{obslog}}
\tablewidth{0pt}
\tablehead{\colhead{Set} & \colhead{Images} & \colhead{Image Names} & \colhead{UT\tablenotemark{a}} & Exp\tablenotemark{b}  & Image Size\tablenotemark{c} }

\startdata

A-L& 1 - 3& IDLTA1IQQ -  IDLTA1ITQ	 &  07:08:16  - 07:32:01 	& 3$\times$348   & 4124$\times$4385	\\  	  	 
A-M & 4 - 6& IDLTA1IVQ -  IDLTA1IXQ	 &  07:35:28  - 07:43:00 	& 3$\times$30 	& 2058$\times$2175\\ 	  	  	 
A-S & 7 - 10 & IDLTA1IYQ -  IDLTA1J1Q	 &  07:47:06  - 07:52:22 	& 4$\times$4 	& 1029$\times$1086\\	 
B-L & 11 - 13 & IDLTA2J3Q - IDLTA2J6Q 	 &  08:42:57  - 09:06:42 	& 3$\times$348 & 4124$\times$4385	\\
B-M & 14 - 16 & IDLTA2J8Q -  IDLTA2JAQ	 &  09:10:10  - 09:17:42 	& 3$\times$30 	& 2058$\times$2175\\
B-S & 17 - 20 & IDLTA2JBQ - IDLTA2JEQ 	 &  09:21:47  - 09:27:03 	& 4$\times$4 	& 1029$\times$1086\\
C-L & 21 - 23 & IDLTA3JGQ -  IDLTA3JJQ	 &  10:18:08  - 10:41:53 	& 3$\times$348 & 4124$\times$4385	\\
C-M & 24 - 26 & IDLTA3JLQ - IDLTA3JNQ 	 &  10:45:20  - 10:52:52 	& 3$\times$30 & 2058$\times$2175 	\\
C-S & 27 - 30 & IDLTA3JOQ - IDLTA3JRQ 	 &  10:56:57  - 11:02:13 	& 4$\times$4 	& 1029$\times$1086\\

\enddata

\tablenotetext{a}{Times of the start and end of each image, on UT 2017 December 17}
\tablenotetext{b}{Number of exposures $\times$ exposure time in seconds}

\tablenotetext{c}{Number of pixels read out from CCD}

\end{deluxetable}

\clearpage

\begin{deluxetable}{lcccccccr}
\tablecaption{Observing Geometry (UT 2017 December 17)
\label{geometry}}
\tablewidth{0pt}
\tablehead{\colhead{UT Date} & \colhead{dRA/dt\tablenotemark{a}} & d$\delta$/dt\tablenotemark{a} & $r_H$\tablenotemark{b} & \colhead{$\Delta$\tablenotemark{c}}  & \colhead{$\alpha$\tablenotemark{d}} & \colhead{$\theta_{- \odot}$\tablenotemark{e}} & \colhead{$\theta_{-V}$\tablenotemark{f}} & \colhead{$\delta_{E}$\tablenotemark{g}}   }

\startdata

07h 00m	  & -1677.03  &  -1473.53 &   1.00557  & 0.06922   & 70.007   & 71.225   & 71.016    & 0.319   \\

08h 00m	& -1407.65   & -1557.29   & 1.00488  & 0.06931   & 70.674   & 70.912   & 70.570   &  0.509	\\

09h 00m	& -2253.04   & -1317.61   & 1.00418  & 0.06932   & 71.255   & 70.631   & 70.117   &  0.755	\\

10h 00m	& -1338.80   & -1579.85   & 1.00348  & 0.06949   & 71.847   & 70.356   & 69.680    & 0.995	\\

11h 00m	 & -1706.39   & -1458.21   & 1.00279  & 0.06955  & 72.506   & 70.070   & 69.273    & 1.186	\\

\enddata


\tablenotetext{a}{Instantaneous rates of angular motion in right ascension and declination, relative to sidereal, in arcsecond hr$^{-1}$}

\tablenotetext{b}{Heliocentric distance, in AU}
\tablenotetext{c}{Geocentric distance, in AU}
\tablenotetext{d}{Phase angle, in degrees}
\tablenotetext{e}{Position angle of projected anti-solar direction, in degrees}
\tablenotetext{f}{Position angle of negative projected orbit vector, in degrees}
\tablenotetext{g}{Angle from orbital plane, in degrees}

\end{deluxetable}

\clearpage

\clearpage

\clearpage

\clearpage

\clearpage

\begin{thebibliography}{}




\bibitem[Ansdell et al.(2014)]{2014ApJ...793...50A} Ansdell, M., Meech, K.~J., Hainaut, O., et al.\ 2014, \apj, 793, 50 

\bibitem[Arai et al.(2018)]{2018LPI....49.2570A} Arai, T., Kobayashi, M., Ishibashi, K., et al.\ 2018, Lunar and Planetary Science Conference, 49, 2570 

\bibitem[Arendt(2014)]{2014AJ....148..135A} Arendt, R.~G.\ 2014, \aj, 148, 135 

\bibitem[Babadzhanov \& Kokhirova(2009)]{2009A&A...495..353B} Babadzhanov, P.~B., \& Kokhirova, G.~I.\ 2009, \aap, 495, 353 

\bibitem[Beech(2002)]{2002MNRAS.336..559B} Beech, M.\ 2002, \mnras, 336, 559 


\bibitem[Blaauw(2017)]{2017P&SS..143...83B} Blaauw, R.~C.\ 2017, \planss, 143, 83 

\bibitem[Bowell et al.(1989)]{1989aste.conf..524B} Bowell, E., Hapke, B., 
Domingue, D., et al.\ 1989, ``Application of photometric models to asteroids'', in Asteroids II, Tucson, AZ, University of Arizona Press, p. 524-556. 


\bibitem[de Le{\'o}n et al.(2010)]{2010A&A...513A..26D} de Le{\'o}n, J., Campins, H., Tsiganis, K., Morbidelli, A., \& Licandro, J.\ 2010, \aap, 513, A26 

\bibitem[DeMario et al.(2016)]{2016Icar..280..308D} DeMario, B.~E., Schmidt, B.~E., Mutchler, M.~J., et al.\ 2016, \icarus, 280, 308 

\bibitem[Hanu{\v{s}}, et al.(2016)]{2016A&A...592A..34H} Hanu{\v{s}}, J., Delbo', M., Vokrouhlick{\'y}, D., et al.\ 2016, \aap, 592, A34.

\bibitem[Hsieh \& Jewitt(2005)]{2005ApJ...624.1093H} Hsieh, H.~H., \& Jewitt, D.\ 2005, \apj, 624, 1093 

\bibitem[Hsieh \& Jewitt(2006)]{2006Sci...312..561H} Hsieh, H.~H., \& Jewitt, D.\ 2006, Science, 312, 561 



\bibitem[Hui \& Li(2017)]{2017AJ....153...23H} Hui, M.-T., \& Li, J.\ 2017, \aj, 153, 23 



\bibitem[Ishibashi et al.(2018)]{2018LPI....49.2126I} Ishibashi, K., Kameda, S., Kagitani, M., et al.\ 2018, Lunar and Planetary Science Conference, 49, 2126 

\bibitem[Ishiguro et al.(2009)]{2009AdSpR..43..875I} Ishiguro, M., Sarugaku, Y., Nishihara, S., et al.\ 2009, Advances in Space Research, 43, 875 

\bibitem[Ito et al.(2018)]{2018NatComm} Ito, T., Ishiguro, M., Arai, T. et al.\ 2018, Nature Communications, 9, 2486

\bibitem[Jakub{\'{\i}}k \& Neslu{\v s}an(2015)]{2015MNRAS.453.1186J} Jakub{\'{\i}}k, M., \& Neslu{\v s}an, L.\ 2015, \mnras, 453, 1186 

\bibitem[Jewitt(2012)]{2012AJ....143...66J} Jewitt, D.\ 2012, \aj, 143, 66 


\bibitem[Jewitt \& Hsieh(2006)]{2006AJ....132.1624J} Jewitt, D., \& Hsieh, H.\ 2006, \aj, 132, 1624 

\bibitem[Jewitt \& Li(2010)]{2010AJ....140.1519J} Jewitt, D., \& Li, J.\ 2010, \aj, 140, 1519 

\bibitem[Jewitt et al.(2013)]{2013ApJ...771L..36J} Jewitt, D., Li, J., \& Agarwal, J.\ 2013, \apjl, 771, L36 

\bibitem[Jewitt et al.(2014)]{2014AJ....147..117J} Jewitt, D., Ishiguro, M., Weaver, H., et al.\ 2014, \aj, 147, 117 

\bibitem[Jewitt et al.(2015)]{2015aste.book..221J} Jewitt, D., Hsieh, H., \& Agarwal, J.\ 2015a, Asteroids IV, Universiy of Arizona Press, Tucson, p. 221 

\bibitem[Jewitt et al.(2015)]{2015ApJ...798..109J} Jewitt, D., Agarwal, J., Weaver, H., Mutchler, M., \& Larson, S.\ 2015b, \apj, 798, 109 

\bibitem[Jewitt et al.(2017)]{2017ApJ...847L..19J} Jewitt, D., Hui, M.-T., Mutchler, M., et al.\ 2017, \apjl, 847, L19 

\bibitem[Kasuga(2009)]{2009EM&P..105..321K} Kasuga, T.\ 2009, Earth Moon and Planets, 105, 321 

\bibitem[Kasuga \& Jewitt(2008)]{2008AJ....136..881K} Kasuga, T., \& Jewitt, D.\ 2008, \aj, 136, 881 

\bibitem[Li \& Jewitt(2013)]{2013AJ....145..154L} Li, J. \& Jewitt, D.\ 2013, \aj, 145, 154.

\bibitem[Madiedo et al.(2013)]{2013MNRAS.436.2818M} Madiedo, J.~M., Trigo-Rodr{\'\i}guez, J.~M., Castro-Tirado, A.~J., et al.\ 2013, \mnras, 436, 2818.

\bibitem[Marcus(2007)]{2007ICQ....29...39M} Marcus, J.~N.\ 2007, International Comet Quarterly, 29, 39 

\bibitem[Masanori et al.(2018)]{2018LPI....49.2050M} Masanori, M., Srama, R., Kr{\"u}ger, H., Arai, T., \& Kimura, H.\ 2018, Lunar and Planetary Science Conference, 49, 2050 

\bibitem[McFadden, et al.(2015)]{2015Icar..257..207M} McFadden, L.~A., Skillman, D.~R., Memarsadeghi, N., et al.\ 2015, \icarus, 257, 207.

\bibitem[Ohtsuka et al.(2006)]{2006A&A...450L..25O} Ohtsuka, K., Sekiguchi, T., Kinoshita, D., et al.\ 2006, \aap, 450, L25 

\bibitem[Ott et al.(2017)]{2017MNRAS.469S.276O} Ott, T., Drolshagen, E., Koschny, D., et al.\ 2017, \mnras, 469, S276 

\bibitem[Pravec et al.(2002)]{2002aste.book..113P} Pravec, P., Harris, A.~W., \& Michalowski, T.\ 2002, Asteroids III, 113 

\bibitem[Rubanenko \& Aharonson(2017)]{2017Icar..296...99R} Rubanenko, L., \& Aharonson, O.\ 2017, \icarus, 296, 99 

\bibitem[Russell(1916)]{1916ApJ....43..173R} Russell, H.~N.\ 1916, \apj, 43, 173 

\bibitem[Ryabova(1999)]{1999SoSyR..33..224R} Ryabova, G.~O.\ 1999, Solar System Research, 33, 224 

\bibitem[Ryabova(2007)]{2007MNRAS.375.1371R} Ryabova, G.~O.\ 2007, \mnras, 375, 1371 





\bibitem[Schult et al.(2018)]{2018Icar..309..177S} Schult, C., Brown, P., Pokorn{\'y}, P., Stober, G., \& Chau, J.~L.\ 2018, \icarus, 309, 177 

\bibitem[Szalay et al.(2018)]{2018MNRAS.474.4225S} Szalay, J.~R., Pokorn{\'y}, P., Jenniskens, P., \& Hor{\'a}nyi, M.\ 2018, \mnras, 474, 4225 

\bibitem[Taylor et al.(2018)]{2018LPI....49.2509T} Taylor, P.~A., Marshall, S.~E., Venditti, F., et al.\ 2018, Lunar and Planetary Science Conference, 49, 2509 

\bibitem[Todorovi{\'c}(2018)]{2018MNRAS.475..601T} Todorovi{\'c}, N.\ 2018, \mnras, 475, 601.

\bibitem[Weaver, et al.(2006)]{2006Natur.439..943W} Weaver, H.~A., Stern, S.~A., Mutchler, M.~J., et al.\ 2006, \nat, 439, 943.

\bibitem[Williams \& Wu(1993)]{1993MNRAS.262..231W} Williams, I.~P., \& Wu, Z.\ 1993, \mnras, 262, 231 


\bibitem[Ye et al.(2018)]{2018ApJ...864L...9Y} Ye, Q., Wiegert, P.~A., \& Hui, M.-T.\ 2018, \apjl, 864, L9 


\end{thebibliography}
\end{document}